\documentstyle[preprint,aps]{revtex} 
\begin{document}
\tightenlines 
\draft


\title{Capillary-gravity wave resistance in ordinary and 
magnetic fluids}

\author{J. Browaeys, 
J.-C. Bacri, R. Perzynski}
\address{Groupe Ferrofluide associ\'e \`a l'Universit\'e Denis Diderot\\ 
Laboratoire des milieux d\'esordonn\'es et h\'et\'erog\`enes, UMR CNRS 7603\\
Universit\'e Pierre et Marie Curie, case 78, 4 place Jussieu 75231 
Paris Cedex 05, France} 
\author{M. I. Shliomis}
\address{Department of Mechanical Engineering, Ben-Gurion University of 
the Negev, P.O.B. 653, Beer-Sheva 84105, Israel}
\date{\today}
\maketitle 
\begin{abstract}
Wave resistance is the drag force associated to the emission of waves 
by a moving disturbance at a fluid free surface.  In the case of 
capillary-gravity waves it undergoes a transition from zero to a 
finite value as the speed of the disturbance is increased.  For the 
first time an experiment is designed in order to obtain the wave 
resistance as a function of speed.  The effect of viscosity is 
explored, and a magnetic fluid is used to extend the available range 
of critical speeds.  The threshold values are in good agreement with 
the proposed theory.  Contrary to the theoretical model, however, the 
measured wave resistance reveals a non monotonic speed dependence 
after the threshold.  
\end{abstract}
%
\vskip1pc 
%

When an object is moved at the free surface of a fluid, it experiences 
a drag force which physically originates from: (a) bulk dissipation 
in a viscous boundary layer for low Reynolds numbers, 
and in the turbulent wake for high Reynolds numbers; (b) the emission of 
capillary-gravity surface waves.  Such waves remove momentum from the 
perturbating object to infinity. The associated force that the 
object experiences is called \emph{wave resistance}.  For the convenient 
moderate speeds on which we focus in this paper it may overcome the 
bulk dissipation type drag.

Wave resistance has been studied for more than a century in the case 
of pure gravity waves \cite{kelvin}, mainly because this topic has 
obvious naval applications\cite{recentgravitywaves}.  In this case the 
variation of the wave resistance $R$ with the speed $V$ follows two 
r\'egimes.  A critical velocity $V_{c}^{grav}$ is imposed by the 
characteristic size $L$ of the ship: $V_{c}^{grav}=\sqrt{gL/2\pi}$, 
$g$ being the gravity acceleration.  For $V<V_{c}^{grav}$ the wave 
resistance is very close to zero and behaves as $R \propto 
\sqrt{V-V_{c}^{grav}}$ for $V>V_{c}^{grav}$.  This has recently been 
analyzed in terms of an imperfect bifurcation by one of the co-authors 
\cite{shliomisprl}.

The case of capillary-gravity waves has been theoretically treated in 
a recent work of Rapha\"el and de Gennes \cite{raphaeldegennes}.  Such 
waves are generated when the size of the perturbating object is small 
compared to the capillary wavelength $\lambda_{c}=2\pi(\sigma/\rho 
g)^{1/2}$, where $\sigma$ is the surface tension of the free air-fluid 
interface and $\rho$ the density of the fluid.  The dispersive 
properties of capillary-gravity waves are such as there exists a 
minimum phase speed $V_{c}=(4\sigma g/\rho)^{1/4}$ at which waves are 
able to propagate.  Since the pattern is stationary in the reference 
frame of the moving object, no wave can be emitted for $V<V_{c}$ 
\cite{lighthill}, and therefore there is no wave resistance in that 
case.  As it has been shown in \cite{raphaeldegennes}, the wave 
resistance experiences a finite jump $R_{c}$ at $V=V_{c}$ and 
increases above $V_{c}$.  The system is thus supposed to undergo a 
discontinuous bifurcation.

In order to check these theoretical predictions it is necessary to 
vary $V_{c}$ by means of $\rho$ and $\sigma$ variations.  By adding a 
surfactant to water, $\sigma$ may be easily chosen between say 20 and 
73 mN/m.  Consequently $V_{c}$ will merely vary from 17 cm/s to 23 
cm/s.  A more efficient control parameter is thus needed.  We here 
show that the action of a magnetic field on a magnetic fluid provides 
a means to tune the critical velocity from its maximum value 
$V_{c}^{H=0}=(4\sigma g/\rho)^{1/4}$ down to 0.  Using a magnetic 
fluid, along with other \emph{regular} fluids of different 
viscosities, we perform $R(V)$ measurements, as the problem has not 
been experimentally treated yet.

\emph{In a regular fluid}, the wave emission process is controlled by 
the dispersion equation of capillary-gravity surface waves, 
$\omega^{2}=gk+\sigma k^{3}/\rho$, where $\omega$ is the circular 
frequency and $k$ the modulus of the wave vector.  The condition for 
stationarity of the wave pattern in the frame of reference of the 
moving object is $\omega= kV\cos{\theta}$ ,where $\theta$ is the angle 
between the speed and wave vectors.  Thus we obtain the following 
equation ($k_{c}=\sqrt{\rho g /\sigma }$ is the capillary wavevector):

\begin{equation}
	\left(\frac{k}{k_{c}}\right)^2
	-2 \left(\frac{V}{V_{c}}\cos{\theta}\right)^2 \left(\frac{k}{k_{c}}\right) 
	+1 = 0 \, ,
    \label{stationarity no field}
\end{equation}
which has no solution for $V<V_{c}$. For a moving Dirac 
Delta pressure distribution $P(x,y,t)=p\delta(x-Vt,y)$ the wave 
resistance is \cite{raphaeldegennes}: 
\begin{equation}
	R=\frac{p^{2}}{\pi \sigma} 
	\int_{0}^{\arccos{\frac{V_{c}}{V}}}\cos{\theta} \ 
	\frac{k_{+}(\theta)^2+k_{-}(\theta)^2}{k_{+}(\theta)-k_{-}(\theta)}\, 
	d\theta \, , 
    \label{resistance general}
\end{equation}
where $k_{+}(\theta)$ and $k_{-}(\theta)$ are the two roots of Eq.  
(\ref{stationarity no field}).  This formula remains valid as long as 
the characteristic size of the pressure distribution in experiments is 
much smaller than the capillary wavelength.  In those conditions, 
close and above the threshold, the wave resistance has a finite value 
$R_{c}=p^{2}k_{c}/2\sqrt{2}\sigma$ and increases monotonically with 
speed (see inset of Fig.  \ref{R(V)gly} and the uppermost curve in 
Fig.  \ref{R(V)FF}).

In order to measure $R(V)$ for the various fluids, they are placed 
into a circular channel dug into a Teflon covered aluminum dish.  The 
latter is fixed to a shaft and rotated at constant rate, thus 
simulating a steady flow for the fluid.  The radius of the channel is 
\mbox{20 cm}, its width is \mbox{2 cm}.  A \mbox{4 cm} wide channel is 
also used, showing no significant difference in the experimental 
results.  The depth of the fluid is usually more than \mbox{1 cm}, 
ensuring the validity of the infinite depth approximation.

The disturbing object consists of a vertical bronze wire (radius 
$r=0.2 \textrm{ mm}$) whose tip just touches the surface of the fluid 
(the wire is wetted by a few tenths of millimeters of fluid).  The 
deflection of the wire is proportional to the horizontal force exerted 
on its free end (which is typically in the order of a micronewton).  
It is measured with an infrared optical sensor.  The calibration of 
the sensor is obtained by tilting the base to which the wire is 
attached.  A more detailed description of the measuring method will be 
published later.

Though no theory includes 3D viscous effects so far, we measure the 
wave resistance for different viscosities.  To this purpose several 
mixtures of water and glycerol are used: the surface tension $\sigma$ 
and the densities $\rho$ of the mixtures are very close to one another 
(see Table 1) so that the impact of viscosity alone may be monitored 
in our experiments.  The viscosities are measured with a standard 
Poiseuille viscometer.

Fig.\ref{R(V)gly} displays the variation of the experimental drag 
$R_{exp}$ as a function of speed for the various mixtures.  All the 
measurements are obtained by increasing and then decreasing the speed: 
there is \emph{no hysteresis}.  We may note that:

(a)\emph{ There is a critical velocity at which the measured drag is 
discontinuous.} That point validates an important feature of the 
Rapha\"el and de Gennes' theory \cite{raphaeldegennes}.  Besides, it 
has been checked that the drag discontinuity occurs at the very speed 
at which the wave pattern develops.  The measured critical velocity 
$V_{c}$ is 23 $\pm$ 0.5 cm/s for pure water.  It corresponds to a 
surface tension interval of [65.1;77.7] mN/m into which lies the 
tabulated value of pure water surface tension 72.75 mN/m at 20 
$^{\circ}$C.  For water/glycerol mixtures we obtained 
\mbox{$V_{c}\approx$ 22.5 cm/s}, that is compatible within 
experimental error bars with the surface tension of the mixtures 
(around \mbox{70 mN/m}).

(b) \emph{The experimental drag is not null below the critical 
velocity}, all the more since the viscosity is high.  The viscous drag 
$R_{drag}$ that is exerted over the immersed wire must be added to the 
wave resistance $R$ to account for the measured drag $R_{exp}$.  At a 
speed of 10 cm/s, the Reynolds number based on the approximate length 
of the wetted part of the wire $h\approx 0.4\ \textrm{mm}$ is already 
equal to 40 (for water).  A crude estimate of the viscous drag can 
still be given by the Stokes formula \cite{landau}: $R_{drag}\approx 
6\pi \eta h V$.  We experimentally check that the viscous drag is 
indeed proportional to $h$; it can be seen in Fig.\ref{R(V)gly} 
(dotted lines) that it is also proportional to $V$, at least for 
moderate enough speeds.  The linear $R_{drag}$ dependence on viscosity 
$\eta$ is harder to assess because it is impossible to impose exactly 
the same $h$ from an experiment to another, the wetting of the wire 
being imperfect; nevertheless it is linear within the uncertainties 
over $h$.  Inset of Fig.\ref{R(V)gly} presents the $R(V)$ variations 
after subtraction of the viscous drag $R_{drag}$ for each sample.  It 
is this quantity that has to be compared with the theoretical 
expression (\ref{resistance general}) (full line in the inset).  It is 
clear from the inset that a pretransitional effect takes place, as the 
measured drag sharply increases just below the threshold (the higher 
the viscosity, the stronger the effect).  A recent model 
\cite{raphael2} for 2D viscous wave resistance predicts such a 
feature.

(c) \emph{The amplitude of the wave resistance discontinuity at 
$V_{c}$ compares well with the theory}.  Assuming a perfect wetting of 
the wire by the fluid, the total force acting on the fluid is $p=2\pi 
r \sigma$ ($r$ is the radius of the wire).  Thus an estimate of the 
wave resistance at the threshold is given by 
$R_c=\pi^{2}r^2\sqrt{2\rho g \sigma}$.  A comparison between expected 
values and what is observed is given in Table \ref{thetable}.  The 
discrepancy is partially due to the imperfect wetting of the fluid on 
the wire, which leads to overestimate the applied vertical force.  On 
the other hand the drag values close to the threshold fluctuate a lot.

(d) \emph{The drag is a non-monotonic function of speed for 
$V>V_{c}$.} In fact, it can be seen in inset of Fig.  \ref{R(V)gly} 
that for $V>V_{c}$ the wave resistance $R$ first \emph{decreases} as 
the speed increases, and then increases again for high enough speeds.  
This feature is not predicted by the current theory, which anyway 
overestimates the actual drag.  Moreover it is unlikely a viscosity 
effect since it is as much marked as the viscosity is low.  It is 
possibly a general feature of such a flow, and in this case the theory 
should be revised to include viscosity and non-linear aspects.


\emph{In a magnetic fluid} the dispersion equation of 
capillary-gravity surface waves is modified with allowance for a 
vertical uniform magnetic field \cite{browaeys}:
\begin{equation}
	\omega^{2}=gk+\frac{\sigma k^{3}}{\rho}
	-\mu_{0}\frac{(\mu_{r}-1)^{2}}{\mu_{r}(\mu_{r}+1)} \frac{H^{2}k^{2}}{\rho}, 
	\label{dispersion w field}
\end{equation}
where $H$ is the magnetic field, $\mu_{r}$ the relative magnetic 
permeability of the magnetic fluid (assumed to be constant 
\cite{browaeys}) and $\mu_{0}$ the vacuum magnetic 
permeability.  For a given wave vector, an increase of the field 
intensity lowers the frequency of the waves.  The frequency drops to 
zero when $H$ reaches a certain critical value $H_{*}$ defined by:
\begin{equation}
H_{*}^{2}=2 \ \frac{\mu(\mu+1)}{(\mu-1)^2}\ \frac{\sqrt{\rho g 
\sigma}}{\mu_{0}}.
\label{Hc}
\end{equation}

For $H>H_{*}$ the surface becomes unstable: the Rosensweig 
instability, sometimes called the peak instability, develops yielding 
an hexagonal array of peaks\cite{rosensweig}.  The condition for 
stationarity implies that $k$ must be a solution of:
\begin{equation}
	\left(\frac{k}{k_{c}}\right) ^2
	-2 \left(\left(\frac{V}{V_{c}}\cos{\theta}\right)^2 + 
	\left(\frac{H}{H_{*}}\right)^2\right) 
	\left(\frac{k}{k_{c}}\right) 
	+1 = 0.
    \label{stationarity w field}
\end{equation}
Real solutions exists if and only if:
\begin{equation}
	V > V_{c}^{H} \textrm{ with }  V_{c}^{H}=V_{c}\sqrt{1-\left(\frac{H}{H_{*}}\right)^2}\, ,
    \label{VCH}
\end{equation}
therefore a steady vertical magnetic field should allow the tuning of the 
critical velocity at which waves (and wave resistance) appear.  

The wave resistance, following Eq.(\ref{resistance general}) and 
Eq.(\ref{stationarity w field}), is given by the following integral:
\begin{equation}
	R^{H}(V)=\frac{p^{2}k_{c}}{\pi \sigma} 
\int_{0}^{\arccos{\frac{V_{c}^{H}}{V}}} 
\!\!\!\!\!\!\!\!\!\!\!\!\!\cos{\theta} \, \left( 
2B(V,\theta)^{\frac{1}{2}} + B(V,\theta)^{-\frac{1}{2}} \right) 
d\theta ,
    \label{RH(V)}
\end{equation}
\begin{equation}
	\textrm{where } B(V,\theta) = \left( \left( \frac{V}{V_{c}} \cos{\theta} \right)^2  
	 + \left( \frac{H}{H_{*}} \right) ^2 \right)^{2} - 1 \, .
    \label{B}
\end{equation}
Just above the threshold, the wave resistance has the finite value :
\begin{equation}
	R_{C}^{H}= 	
	\frac{p^{2}k_{c}}{2\sqrt{2}\sigma}
	\left( 1-\left(\frac{H}{H_{*}}\right)^2\right)^{-\frac{1}{2}}.
    \label{RCH}
\end{equation}

Another experiment is conducted using a water based magnetic fluid 
synthesized according to the Massart method \cite{massart}. Its 
critical field $H_{*}$ is 9.15 kA/m and its surface tension of 60 mN/m 
doesn't depend on the magnetic field.  Other caracteristics are given in 
table \ref{thetable}. The experimental critical values $V_{c}^{H}$ and 
$R_{c}^{H}=R_{exp}(V_{c}^{H})-R_{drag}(V_{c}^{H})$ are both plotted 
versus the normalized magnetic field $H/H_{*}$ in Fig.  \ref{scalingV} 
and Fig.  \ref{scalingR}, and are compared to theoretical predictions 
(\ref{VCH}) and (\ref{RCH}).

The theoretical expression (\ref{VCH}) of $V_{c}^{H}$ (Fig.  
\ref{scalingV}) remarkably fits the data points --- note that there is 
no adjustable parameters.  A data point lies outside the curve, but 
this is probably related to an imperfect magnetic wetting phenomenon.  
As the magnetic field gets closer to the peak instability threshold 
value $H_{*}$, the fluid "climbs" onto the wire, producing a much 
higher viscous drag, a situation which gets away from our inviscid 
linear theoretical analysis.  This also explains the discrepancy in 
Fig.  \ref{scalingR} between experimental and theoretical $R_{C}^{H}$ 
values.  We do not account for the force that the magnetic field is 
exerting at the meniscus close to the wire.  Indeed, the very shape of 
the meniscus creates a non homogeneous magnetic field which results in 
a force that sucks the magnetic fluid up and changes the shape of the 
meniscus.  Only advanced numerical simulations would allow to compute 
the net force added\cite{drop}.

Fig.  \ref{R(V)FF} presents the results obtained for different 
magnetic fields in a reduced representation 
$R/R_{c}^{H}=f(V/V_{c}^{H})$ with $R^{H}=R_{exp}-R_{drag}$.  It also 
gives a comparison to the theoretical predictions of Eq.(\ref{RH(V)}).  
As it was pointed out with regular viscous fluids, the theoretical 
variations of $R/R_{c}^{H}$ lie above the data points, except for $H 
\approx H_{*}\ $.  Then the experimental data and the theory are very 
comparable.  The present theoretical description thus gives a correct 
general trend for the influence of the field on the wave resistance.

In conclusion, for the first time a capillary-gravity wave resistance 
measurement is performed on fluids of various viscosities.  A drag 
discontinuity is always observed for a critical velocity $V_{c}$.  
Thanks to a magnetic fluid the critical velocity range is 
experimentally extended.  In all cases the measured critical 
velocities and the critical values of the resistance are in good 
accordance with the developped models.  If an inviscid theory is 
correct at the threshold, there are some discrepancies for $V>V_{c}$ 
such as a non-monotonic behavior of the wave resistance.  Viscosity 
and non linear aspects should be taken into account in further works.  
Finally, in order to get rid of the viscous drag that is always 
present in our experiments, another mode of disturbance is to be 
envisaged, such as a small magnet placed just above the free surface 
of a flowing magnetic fluid.

We wish to thank J. Servais and P. Lepert for their technical 
assistance, S. Neveu for providing us with the ferrofluid sample and 
E.  Rapha\"el for helpful comments.


\begin{figure} 
\caption{ Experimental 
drag $R_{exp}$ as a function of speed $V$ for different water glycerol 
mixtures.  For readability purpose only a few error bars are plotted.  
Dotted lines : linear viscous drag $R_{drag}$ for each viscosity.  
Inset : wave resistance $R=R_{exp}-R_{drag}$ as a function of $V$.  
Same symbols as in the main figure.  Full line : theoretical 
expression from Eq.  (2).}
\label{R(V)gly}
\end{figure}
 
\begin{figure}
\caption{Reduced critical speed 
$V_{c}^{H}/V_{c}^{H=0}$ at which wave resistance appears in function 
of the applied reduced magnetic field $H/H_{*}$.  The straight line 
represents the theoretical law given by Eq.(\ref{VCH}).  There is no 
adjustable parameter.}
\label{scalingV}
\end{figure}
 
\begin{figure}
\caption{Drag at threshold $R_{c}^{H}$ 
as a function of the reduced magnetic field $H/H_{*}$.  The full line 
represents the theoretical law as given by Eq.(\ref{RCH}).  There is 
no adjustable parameter.}
\label{scalingR}
\end{figure}
 
\begin{figure}
\vskip5mm \caption{Measured wave resistance 
$R^{H}=R_{exp}^{H}-R_{drag}^{H}$ as a function of reduced speed 
$V/V_{c}^{H}$ for different reduced magnetic fields $H/H_{*}$.  The 
theoretical curves are derived from Eq.(\ref{RH(V)},\ref{B}).  The 
uppermost curve describes the wave resistance of a regular 
non-magnetic fluid.}
\label{R(V)FF}
\end{figure}

\begin{table}
\caption{Experimental drag discontinuity at the threshold compared to 
the theoretical predictions of [4], for various water glycerol 
mixtures and an aqueous magnetic fluid (MF).}
\begin{tabular}{lddddd}
Glycerol mass fraction (\%) & 60 & 44.5 & 30 & 0 & MF \\
Viscosity (mPa.s) & 12.5 & 5.1 & 2.6 & 1.0 & 7.0 \\
Density (g/cm$^{3}$) & 1.16 & 1.13 & 1.09 & 1.00 & 1.56 \\
\tableline 
Theory ($\mu$N) & 3.9 & 3.8 & 3.8 & 3.7 & 4.2\\
Experience ($\mu$N) & 2.9 & 2.6 & 4.0 & 3.6 & 4.0\\ 
Uncertainty ($\mu$N) & 0.3 & 0.3 & 0.4 & 1.8 & 1.0\\ 
\end{tabular}
\label{thetable}
\end{table}


\begin{references}
\bibitem{kelvin} Lord Kelvin, Proc. R. Soc. London A \textbf{42} , 80 
(1887).
\bibitem{recentgravitywaves} A.A Kostyukov, \textit{Theory of Ship Waves and Wave Resistance} (Effective Commun. Inc., Iowa City, 1968).
\bibitem{shliomisprl} M. I. Shliomis and V. Steinberg, Phys. Rev. Lett. \textbf{79}, 4178 (1997).
\bibitem{raphaeldegennes} E. Rapha\"el and P.-G. de Gennes, Phys. Rev. E, \textbf{53} 3448 (1996).
\bibitem{lighthill} J. Lighthill, \textit{Waves ind Fluids}, (Cambridge University Press, Cambridge, 1996).
\bibitem{rosensweig} M. D. Cowley and R. E. Rosensweig, J. Fluid Mech. \textbf{30}, 671 (1967).
\bibitem{landau} L.  D.  Landau and E.  M.  Lifshitz, \textit{Fluid Mechanics}, 2nd ed. (Pergamon Press, New York, 1987).
\bibitem{raphael2} D. Richard and E. Rapha\"el, \textit{to be published} in Europhys.  Lett..  
\bibitem{massart} R. Massart, IEEE Trans.  Magn.  \textbf{17}, 1247 (1981).  
\bibitem{drop} A. G. Boudouvis and L. E. Scriven, J. Magn. Magn. Mater. \textbf{122}, 254 (1993).
\bibitem{browaeys} J. Browaeys, J.-C. Bacri, C. Flament, S. Neveu and R. Perzynski, Eur. Phys. J. B \textbf{9}, 335 (1999).
\end{references}
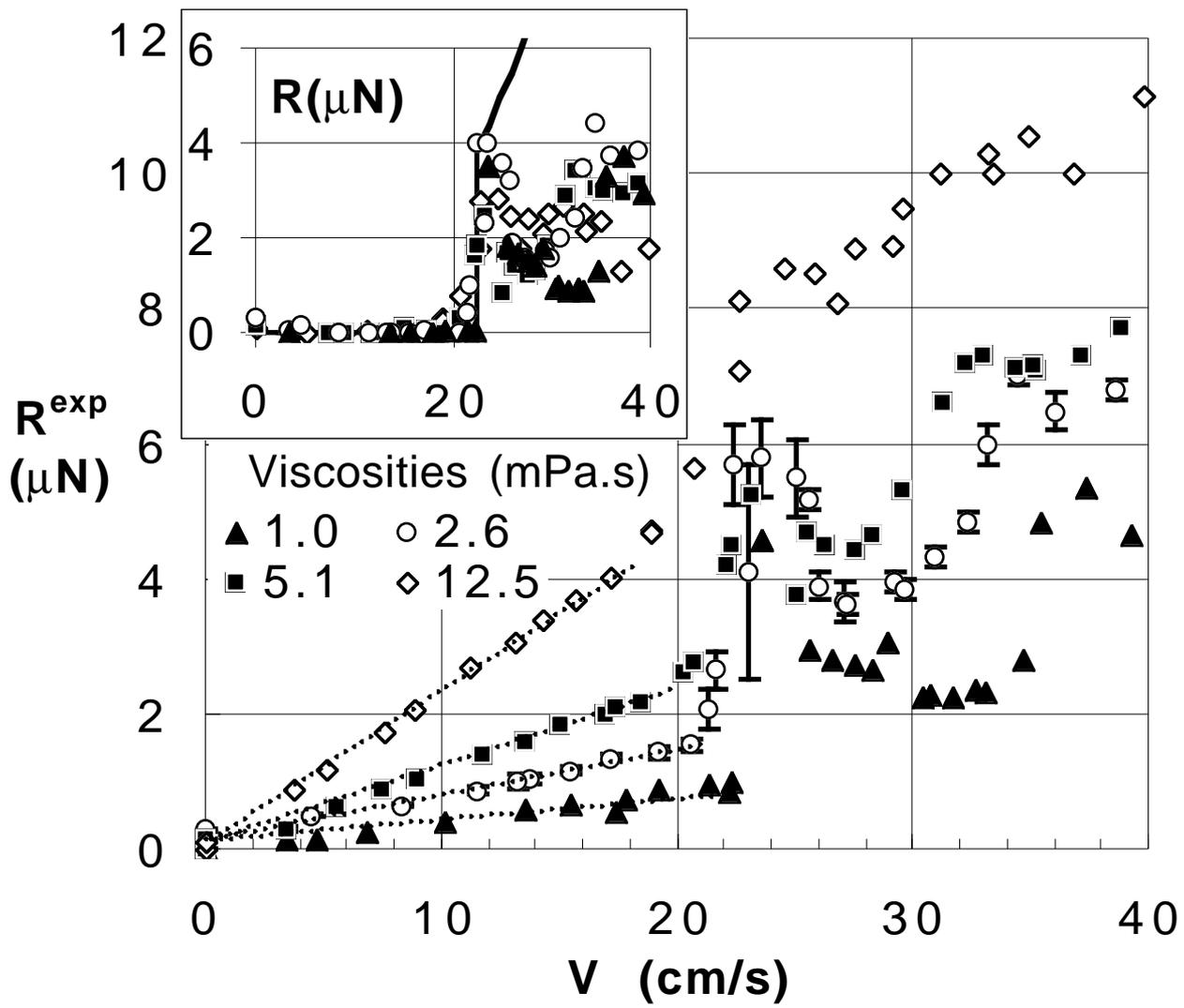
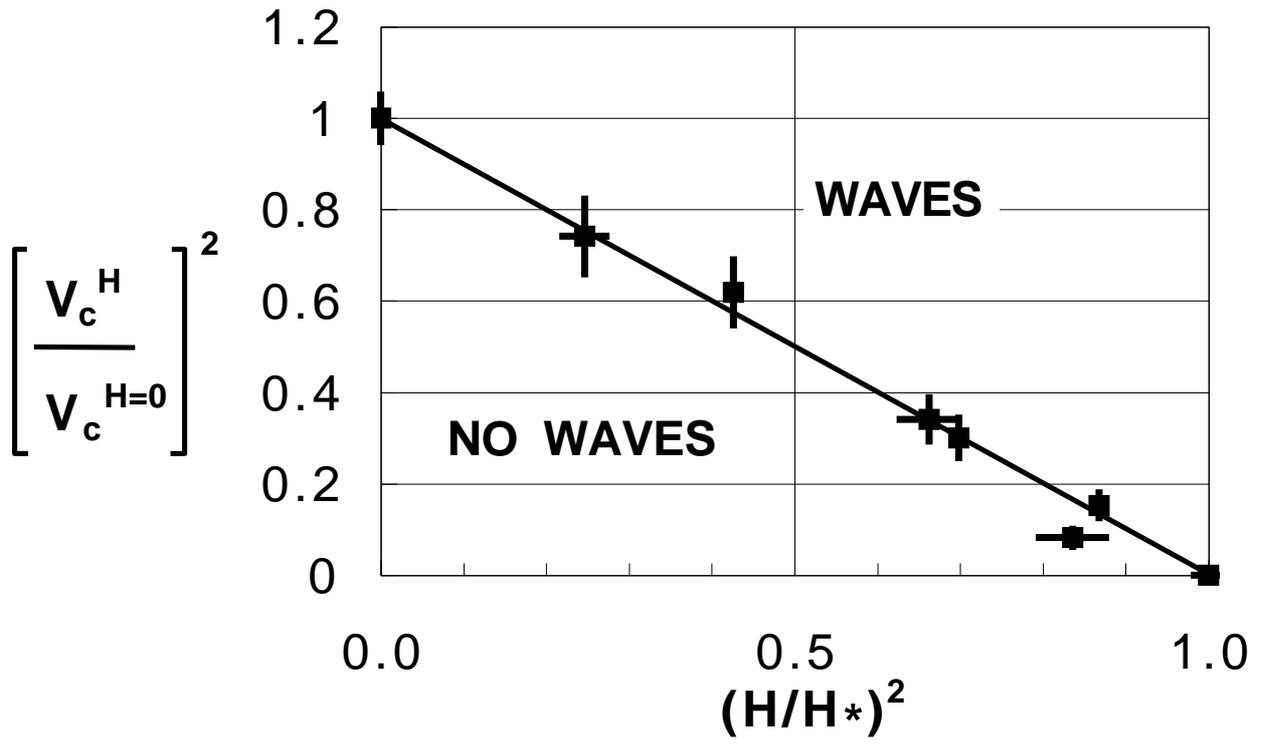
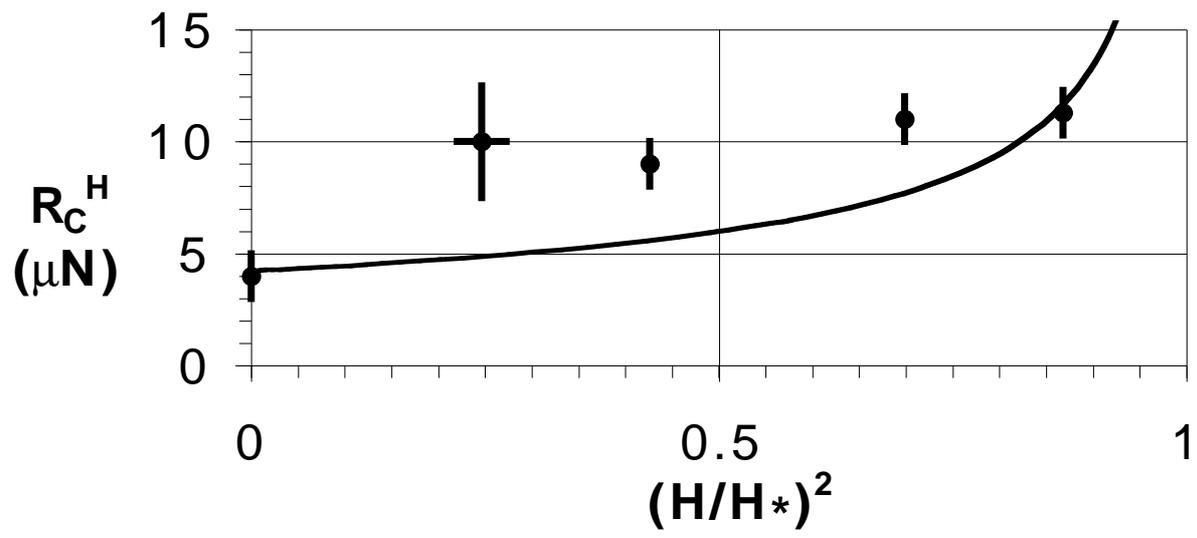
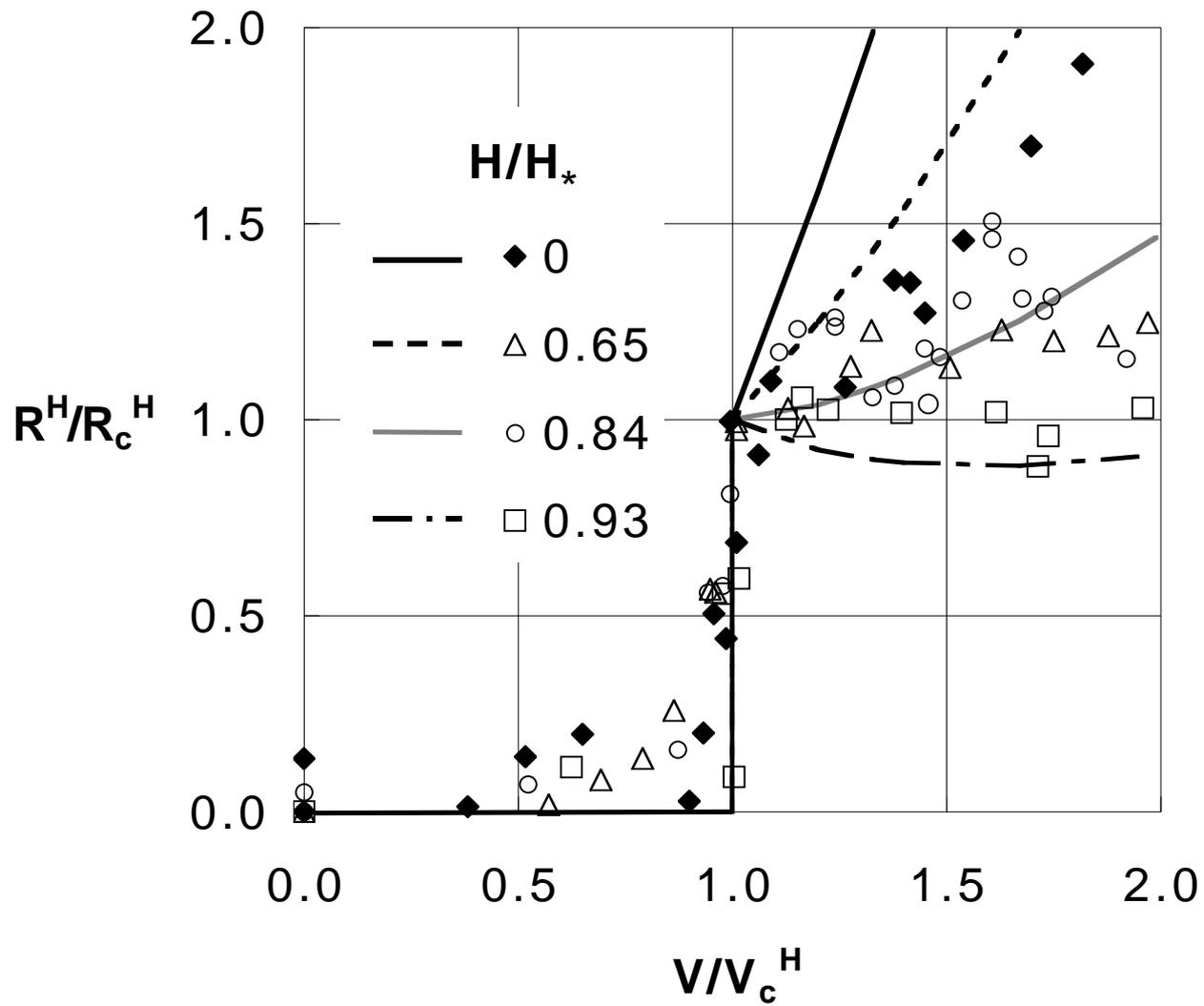
\end{document}